\documentclass[prd,tightenlines,eqsecnum,amsmath,amssymb,nofootinbib,12pt]{revtex4}

\usepackage{epsfig}
\usepackage{bm}% bold math
\begin{document}
\title{Anthropic Reasoning and Quantum Cosmology\footnote{This is a
slightly abbreviated version of an article that appeared in
 {\it The New Cosmology: Proceedings of the Conference on Strings and
Cosmology, College Station, Texas, March 14-17, 2004}, edited by R.
Allen, D. Nanopoulos and C. Pope, AIP Conference Proceedings, v. 743 (American
Institute of Physics, Melville, NY, 2004)}}

\author{James B.~Hartle}
\email{hartle@physics.ucsb.edu}

\affiliation{Department of Physics
 University of California\\
 Santa Barbara, CA 93106-9530 USA}

\date{\today}

\begin{abstract}

Prediction in quantum cosmology requires a specification of the universe's
quantum dynamics {\it and} its quantum state.  We expect only a few general
features of the universe to be predicted with probabilities near unity
conditioned on the dynamics and quantum state alone.  Most useful
predictions are of conditional probabilities that assume additional 
information beyond
the dynamics and quantum state.  Anthropic reasoning utilizes 
probabilities conditioned on `us'. This paper discusses the utility,
limitations, and theoretical uncertainty involved in using such
probabilities.  The predictions resulting from various levels of ignorance
of the quantum state are discussed.

\end{abstract}

%\pacs{}
\maketitle

%==========BODY OF PAPER =========================================

%\setlength{\baselineskip}{.3in}

\section{Introduction}

If the universe is a quantum mechanical system, then it has a quantum
state.  This state provides the initial condition for cosmology. A theory of
this state is an essential part of any final theory summarizing the
regularities exhibited universally by all physical systems and is the
objective of the subject of quantum cosmology.  This essay is concerned
with the role the state of the universe plays in anthropic reasoning --- the
process of explaining features of our universe from our existence in it
\cite{BT86}. The thesis will be that anthropic reasoning in a quantum
mechanical context depends crucially on assumptions about the universe's 
quantum state.

\section{A Model Quantum Universe}

Every prediction in a quantum mechanical universe depends on its state if
only very weakly. Quantum mechanics predicts probabilities
for alternative possibilities, most generally the probabilities for
alternative histories of the universe. The computation of these 
probabilities requires both a theory of the quantum state as well as the 
theory of the dynamics specifying its evolution.

To make this idea concrete while keeping the discussion manageable, we
consider a model quantum universe. The details of this model are not
essential to the subsequent discussion of anthropic reasoning but help to
fix the notation for probabilities and provide a specific example of what
they mean.
Particles and fields move in a large, perhaps expanding box, say presently
20,000 Mpc on a side.  Quantum gravity is neglected --- an excellent
approximation for accessible alternatives in our universe later than
$10^{-43}$s from the big bang.
Spacetime geometry is thus fixed with a well defined notion of time and the
usual quantum apparatus of Hilbert space, states, and their unitary
evolution governed by a Hamiltonian can be applied\footnote{For a more
detailed discussion of this model in the notation used here, see
\cite{Har93a}. For a quantum framework when spacetime geometry is not
fixed, see {\it e.g.}~\cite{Har95c}.}. 

The Hamiltonian $H$ and the state $|\Psi\rangle$ in the Heisenberg picture
are the assumed theoretical inputs to the prediction of quantum mechanical
probabilities.
Alternative possibilities at one moment of time $t$ can be reduced to
yes/no alternatives represented by an exhaustive set of orthogonal projection 
operators
$\{P_\alpha (t)\}$, $\alpha=1, 2, \cdots$ in this Heisenberg picture.  The
operators representing the same alternatives at different times are
connected by
\begin{equation}
P_\alpha(t) = e^{iHt/\hbar} P_\alpha(0)\, e^{-iHt/\hbar}\, .
\label{twoone}
\end{equation}
For instance, the $P$'s could be projections onto an exhaustive  set of exclusive ranges
of the center-of-mass position of the Earth labeled by $\alpha$. The
probabilities $p(\alpha)$ that the Earth is located in one or another of
these regions at time $t$ is
\begin{equation}
p\,(\alpha|H, \Psi) = \Vert P_\alpha (t)\, |\Psi\rangle\Vert^2\, .
\label{twotwo}
\end{equation} 
The probabilities for the Earth's location at a different time is given by
the same formula with different $P$'s computed from the Hamiltonian by
\eqref{twoone}. The notation $p\, (\alpha|H, \Psi)$ departs from usual
conventions (e.g. \cite{Har93a}) to indicate explicitly that all probabilities are conditioned
on the theory of the Hamiltonian $H$ and quantum state $|\Psi\rangle$.

Most generally quantum theory predicts the probabilities of sequences of
alternatives at a series of times --- that is {\it histories}. An example is 
a sequence of ranges of center of mass position of
the Earth at a series of times giving a coarse-grained description of its
orbit. Sequences of sets of alternatives
$\{P^k_{\alpha_k} (t_k)\}$ at a series of times $t_k$, $k=1, \cdots, n$
specify a set of alternative  histories of the model universe. An
individual history $\alpha$ in the set corresponds to a particular sequence
of alternatives $\alpha \equiv(\alpha_1, \alpha_2, \cdots, \alpha_n)$ and is
represented by the corresponding chain of projection operators $C_\alpha$
\begin{equation}
C_\alpha\equiv P^n_{\alpha_n} (t_n) \cdots P^1_{\alpha_1} (t_1) \quad ,
\quad \alpha\equiv (\alpha_1, \cdots, \alpha_n)\, .
\label{twothree}
\end{equation}
The probabilities of the histories in the set are given by
\begin{equation}
p\, (\alpha|H, \Psi) \equiv p\, (\alpha_n,\cdots ,\alpha_1|H, \Psi)= \Vert
C_\alpha|\Psi\rangle\Vert^2
\label{twofour}
\end{equation}
{\it provided} the set decoheres, {\it i.e.}~provided the branch state
vectors $C_\alpha|\Psi\rangle$ are mutually orthogonal. Decoherence ensures
the consistency of the probabilities \eqref{twofour} with the usual rules of
probability theory\footnote{For a short introduction to decoherence see
 \cite{Har93a} or any of the classic
expositions of decoherent (consistent) histories quantum theory \cite{Gri02,
Omn94, Gel94}.}.
 
To use either \eqref{twotwo} or \eqref{twofour} to make predictions, a theory of {\it both} $H$ and $|\Psi\rangle$ is needed. No state; no predictions. 

\section{What is Predicted?}

``If you know the wave function of the universe, why aren't you rich?''
This question was once put to me by my colleague Murray Gell-Mann.  The
answer is that there are unlikely to be any alternatives relevant to making
money that are predicted as sure bets conditioned just on the
Hamiltonian and quantum state alone.  A probability $p({\rm rise}|H,\Psi)$ for the stock market to
rise tomorrow could be predicted from $H$ and $|\Psi\rangle$ through \eqref{twotwo}  in principle.  But
it seems likely that the result would be a useless $p({\rm
rise}|H,\Psi)\approx 1/2$ 
conditioned
just on the `no boundary' wave function \cite{Haw84a} and M-theory.

It's plausible that this is the generic situation.  To be manageable and
discoverable, the theories of dynamics and the quantum state must be short
--- describable we  in terms of a few fundamental equations and the
explanations of the symbols they contain. It's therefore unlikely that
$H$ and $|\Psi\rangle$ 
contain enough information to determine most of the interesting complexity of
the present universe with significant probability \cite{Har96b,Har03}. We
{\it hope} that the Hamiltonian and the quantum state are sufficient conditions
to predict certain large scale features of the
universe with significant probability. 
Approximately classical spacetime, the number of large spatial
dimensions, the approximate homogeneity and isotropy on scales above
several hundred Mpc, and the spectrum of density fluctuations that
were the input to inflation are some examples of these. But even a simple
feature like the time the Sun will rise tomorrow will
not be usefully predicted by our present theories of dynamics and the
quantum state {\it alone}.

The time of sunrise {\it does} become predictable with high probability if a few
previous positions and orientations of the Earth in its orbit are supplied
in addition to $H$ and $|\Psi\rangle$.
That is a particular case of a {\it conditional probability} of the form
\begin{equation}
p\, (\alpha|\beta, H, \Psi) = \frac{p\, (\alpha, \beta | H, \Psi)}{p\,
(\beta | H, \Psi)}
\label{threeone}
\end{equation}
for alternatives $\alpha$ ({\it e.g.}~the times of sunrise) given $H,
|\Psi\rangle$
and further alternatives $\beta$ ({\it e.g.}~a few earlier positions and
orientations of the Earth). The joint probabilities on the right hand side of
\eqref{threeone} are computed using \eqref{twofour} as described in Section II.

Conditioning probabilities on specific information can weaken their
dependence on $H$ and $|\Psi\rangle$ but does not eliminate it.  That is
because any specific information available to us as human observers 
(like a few positions of
the Earth) is but a small part of that needed to specify the state of the
universe.  The $P_\beta$ used to define the joint probabilities in
\eqref{threeone} by 
\eqref{twofour} therefore spans a very large subspace of Hilbert space. 
As a consequence  $P_\beta |\Psi \rangle$  depends strongly on
$|\Psi\rangle$. For example, to
extrapolate present data on the Earth to its position 24 hours from now
requires that the probability be high that it moves on a classical orbit in
that time and that the probability be low that it is destroyed by a neutron
star now racing across the galaxy at near light speed.  Both of these
probabilities depend crucially, if weakly, on the nature of the quantum
state \cite{Har94b}. 

Many useful predictions in physics are of conditional probabilities of the
kind discussed in this section. We next turn to the question of whether we
should be part of the conditions.

\section{Anthropic Reasoning --- Less is More}

\subsection{Anthropic Probabilities}

In calculating the conditional probabilities for predicting some of {\it
our} observations given others, there can be no objection of principle to
including a description of `us' as part of the conditions,
\begin{equation}
p\, (\alpha|\beta, \textrm{`us'}, H, \Psi)\, .
\label{fourone}
\end{equation} 
Drawing inferences using such probabilities is called {\it anthropic
reasoning}. The motivation is the idea is that probabilities for certain 
features of the
universe might be sensitive to this inclusion.

The utility of anthropic reasoning depends on how sensitive probabilities like \eqref{fourone}
are to the inclusion of `us'. To make this concrete, consider the
probabilities for a hypothetical cosmological parameter we will call 
$\Lambda$. We will
assume that $H$ and $|\Psi\rangle$ imply that $\Lambda$ is constant over
the visible universe, but only supply probabilities for the various constant values
it might take through \eqref{twofour}.  
We seek to compare $p\, (\Lambda|H, \Psi)$ with $p\,
(\Lambda|\textrm{`us'}, H, \Psi)$.  In principle, {\it both} are
calulable from \eqref{twofour} and \eqref{threeone}. Figure 1 shows three 
possible ways they might be related:

\begin{itemize}

\item $p\, (\Lambda|H, \Psi)$ is peaked around one value as in Fig.~1(a).
The parameter $\Lambda$ is determined either by $H$ or $|\Psi\rangle$, or
by both.\footnote{As, for example, in the as yet inconclusive discussions of
baby universes \cite{Haw84b}.} Anthropic reasoning is not necessary;
the parameter is already determined by fundamental physics.

\item $p\, (\Lambda|H, \Psi)$ is distributed and $p\, (\Lambda |
\textrm{`us'}, H, \Psi)$ is also distributed as in Fig.~1(b). Anthropic
reasoning is inconclusive. One might as well measure the value of $\Lambda$
and use this as a condition for making further predictions\footnote{As
stressed by Hawking and Hertog \cite{HH04}.} {\it i.e.} work with
probabilities of the form $p\, (\alpha|\Lambda, H, \Psi)$.

\item $p\, (\Lambda|H,\Psi)$ is distributed but $p\, (\Lambda
|\textrm{`us'}, H, \Psi)$ is peaked. Anthropic reasoning helps to explain
the value of $\Lambda$.

\end{itemize}

\begin{figure}[t]
\epsfig{file=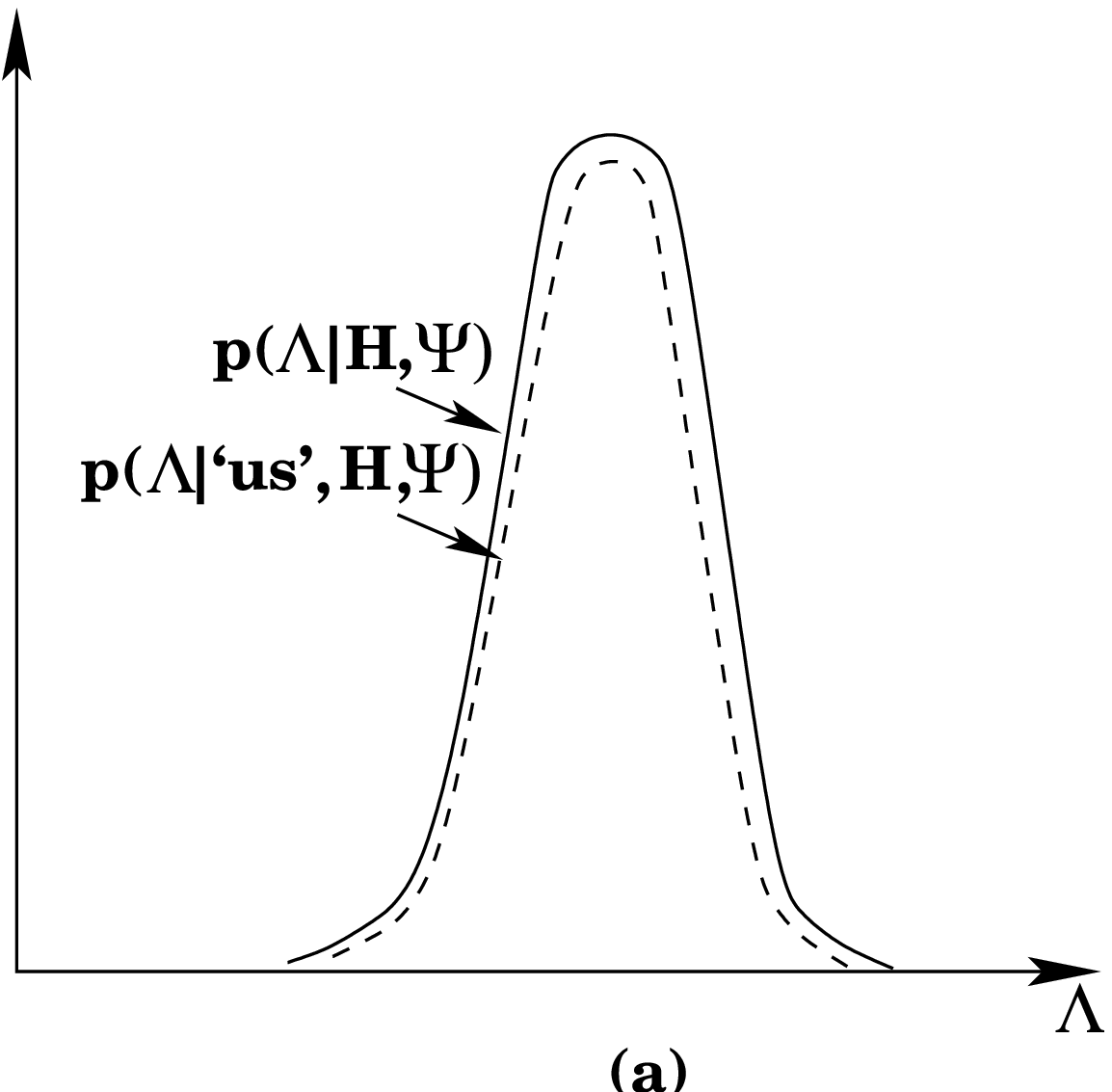, width=2.00in}
\epsfig{file=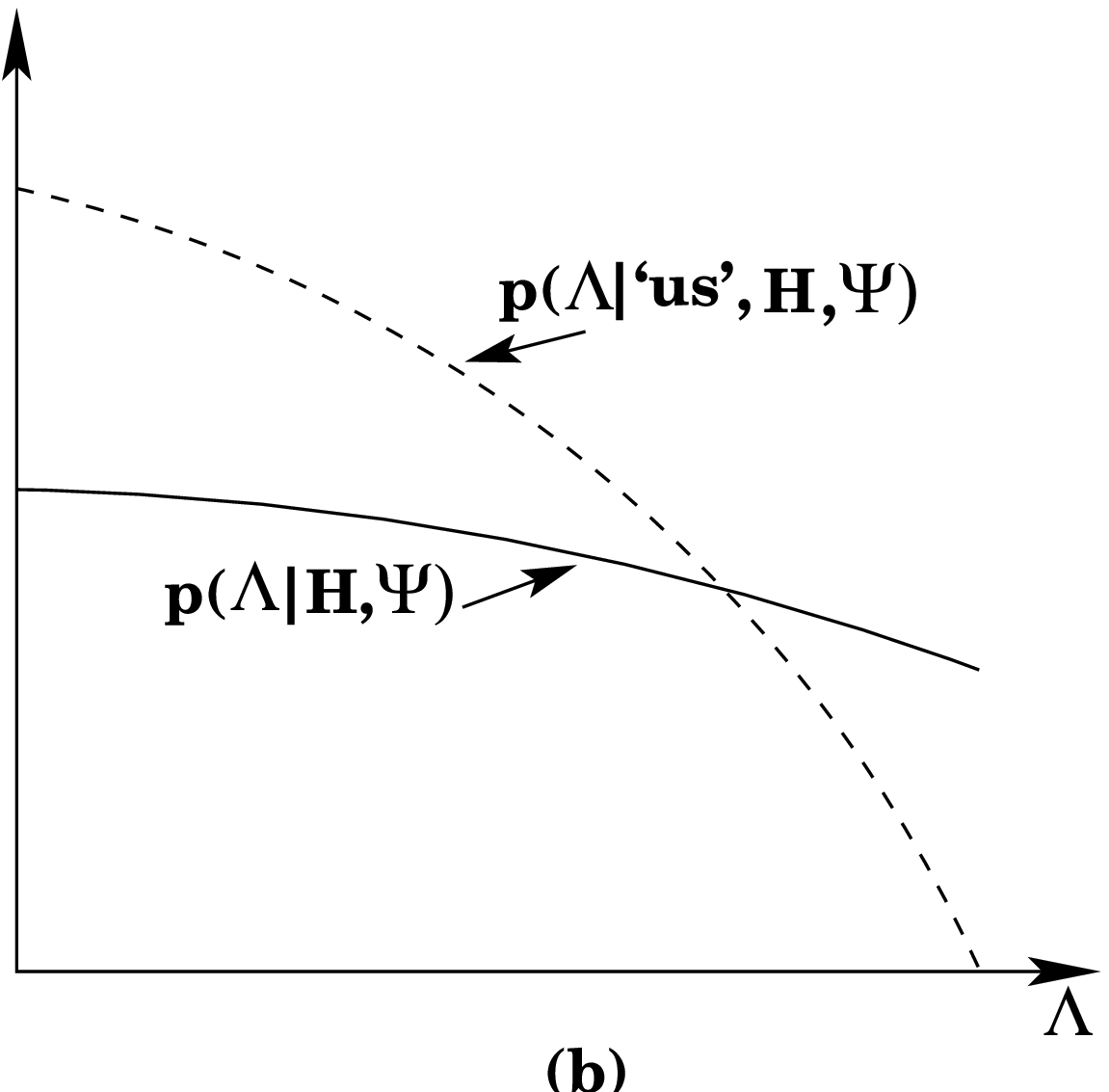, width=2.00in}
\epsfig{file=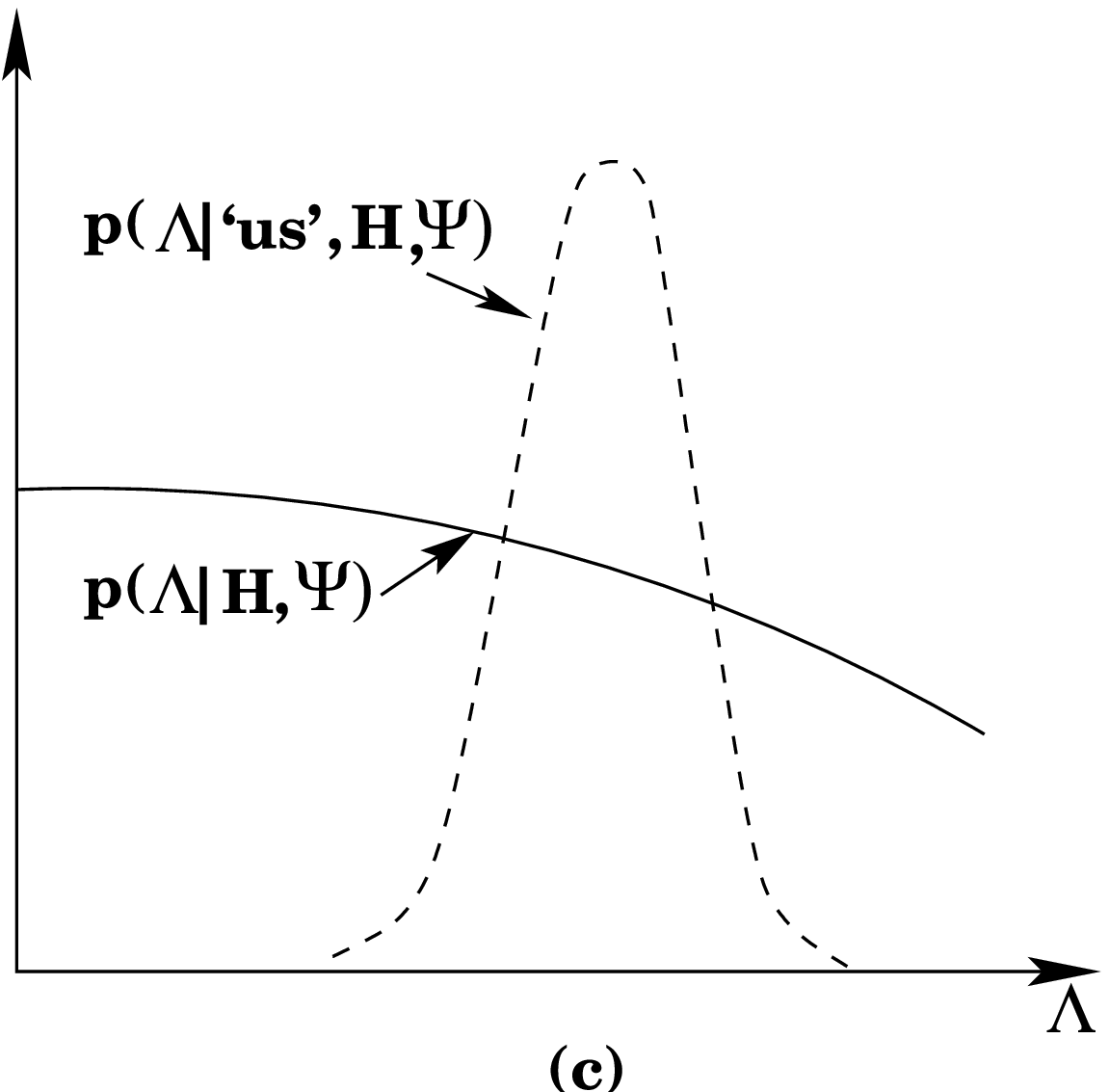, width=2.00in}
\caption{Some possible behaviors for probabilities for the value of a
cosmological parameter $\Lambda$ with and without the condition `us' are illustrated.
In the situation illustrated in (a) the value of $\Lambda$ is fixed by $H$
and $|\Psi\rangle$ and anthropic reasoning is not needed.  In (b) anthropic
probabilities are distributed so that anthropic reasoning is useless in
fixing $\Lambda$. Anthropic reasoning is useful in the situation (c).}
\end{figure}

The important point to emphasize is that a {\it
theoretical hypothesis for} $H$ {\it and} $|\Psi\rangle$ {\it is needed to
carry out anthropic reasoning.} Put differently, a theoretical context is
needed to decide whether a parameter like $\Lambda$ {\it can} vary, and
to find out {\it how} it varies,  before
using anthropic reasoning to restrict its range.  The Hamiltonian 
and quantum
state provide this context. In the Section V  we will consider the
situation where the state is imperfectly known. 

\subsection{Less is More}

While there can be no objections of principle to including `us' as a condition
for the probabilities of our observations, there are formidable obstacles of
practice:

\begin{itemize}

\item We are complex physical systems requiring an extensive 
environment and a long evolutionary history whose description in terms 
of the fundamental variables of $H$
and $|\Psi\rangle$ may be uncertain, long, and complicated.

\item The complexity of the description of a condition including `us' may
make the calculation of the probabilities long or impossible as a practical
matter. 

\end{itemize}
In practice, therefore, anthropic probabilities \eqref{fourone} can only 
be estimated
or guessed. Theoretical uncertainty in the results is thereby introduced.

The objectivity striven for in physics consists, at least in part, in
using probabilities that are not too sensitive to `us'. We would not have
science if anthropic probabilities for observation depended significantly
on which individual human being was part of the conditions.  The existence
of schizophrenic delusions shows that this is possible so that the notion
of `us' should be restricted to exclude such cases. 

For these reasons it is prudent to condition probabilities, not on a
detailed description of `us', but on the weakest condition consistent with
`us' that plausibly provides useful results like those illustrated in
Fig. 1c. A short list of conditions of 
roughly decreasing complexity might include:

\begin{itemize}

\item human beings;
\item carbon-based life;
\item information gathering and utilizing systems (IGUSes);
\item at least one galaxy;
\item a universe older than 5 Gyr;
\item no condition at all.

\end{itemize}
For example, the probabilities used to bound the cosmological constant
$\Lambda$ \cite{BT86, WV01} make use of the fourth and fifth 
on this list under the assumption that including earlier ones will not much
affect the anthropically-allowed range for $\Lambda$.
To move down in the above list of conditions is to move in the direction of
increasing theoretical certainty and decreasing computational complexity.
With anthropic reasoning, less is more.

\section{Ignorance is NOT Bliss}

The quantum state of a single isolated subsystem generally cannot be
determined from a measurement carried out on it.  That is because the
outcomes of measurements are distributed probabilistically and the outcome
of a single trial does not determine the distribution.  Neither can the
state be determined from a series of measurements because measurements 
disturb the state of the
subsystem.  The Hamiltonian can not be inferred from a sequence of
measurements on one subsystem for similar reasons. In the same way, we
can not generally determine either the Hamiltonian or the quantum state of
the universe from our observations of it.  Rather these two parts of a
final theory are theoretical proposals, inferred from partial data to be sure, but
incorporating theoretical assumptions of simplicity, beauty, coherence,
mathematical precision, etc. To test these proposals we search among the
conditional probabilities they imply for predictions of observations yet to
be made with probabilities very near one.  When such predictions occur we
count it a success of the theory, when they do not we reject it and propose
another.

Do we need a theory of the quantum state? To analyze this question, let us
consider various degrees of theoretical uncertainty about it. 

\subsection{Total Ignorance}

In the model cosmology in a box of Section II, 
theoretical uncertainty about the quantum state can be represented by a
density matrix $\rho$ that specifies probabilities for its  eigenstates to be $|\Psi\rangle$.
Total ignorance of the quantum state is represented by a $\rho$ proportional
to the unit matrix.
To illustrate this and the subsequent discussion, assume for the moment
that the dimension of the Hilbert space is very large but finite.  Then
total ignorance of the quantum state is represented by
\begin{equation}
\rho_{\rm tot.~ign.} = \frac{I}{Tr(I)}
\label{fiveone}
\end{equation}
which assigns equal probability any to any member of any complete set of
orthogonal states.

The density matrix \eqref{fiveone} predicts thermal equilibrium, infinite
temperature, infinitely large field fluctuations, and maximum entropy \cite{Har03}.  In
short, its predictions are inconsistent with observations.  This is a
more precise way of saying that every useful prediction depends in some way
on a theory of the quantum state.  Ignorance is not bliss.

\subsection{What We Know}

A more refined approach to avoiding theories of the quantum state is to
assume that it is unknown except for reproducing our present observations
of the universe.  The relevant density matrix is 
\begin{equation}
\rho_{\rm obs} = \frac{P_{\rm obs}}{Tr (P_{\rm obs})}
\label{fivetwo}
\end{equation}          
where $P_{\rm obs}$ is the projection on our current observations ---
``what we know''.
`Observations' in this context mean what we directly observe and record here on
Earth and not the inferences we draw from this data about the larger universe. 
That is because those inferences are based on assumptions about the 
very quantum state that \eqref{fivetwo} aims to ignore. For instance,
we observed nebulae long before we understood what they were or where
they are. The inference that the nebulae are distant clusters of stars and gas 
relies on assumptions about how the
universe is structured on very large scales that are in effect weak 
assumptions on the quantum state. 

Even if we made the overly generous assumption that we had somehow directly 
observed and recorded every detail of the volume 1 km above the surface of 
the Earth, say 
at a 1 mm resolution, that is still a tiny fraction $(\sim 10^{-60})$ of 
the volume inside the present cosmological horizon.  The projection operator 
$P_{\rm obs}$ therefore defines a very large subspace of Hilbert space.  We can
expect that the entropy of the density matrix \eqref{fivetwo} will
therefore be near
maximal, close to that of \eqref{fiveone}, and its predictions similarly
inconsistent with further observations.

In the context of anthropic reasoning, these results show that
conditioning probabilities on `us' alone is not enough to make useful
predictions.  Rather, a theory of $H$ and $|\Psi\rangle$ are needed 
in addition as
described in the previous section.

\section{A Final Theory}

Let us hope that one day we will have a unified theory based on a 
{\it principle} that will specify {\it both} quantum dynamics $(H)$ and a 
unique quantum state of the universe ($|\Psi\rangle$). That would truly be 
a final theory and a proper context for anthropic reasoning.

\acknowledgments

Appreciation is expressed to the Mitchell Institute for hospitality and to
the National Science Foundation for partial support under grant
NSF-PHY02-44764.

\end{document}